%% file: wz_2010_PLB.tex
\RequirePackage{lineno}
\documentclass[aps,prl,twocolumn,showpacs,superscriptaddress,groupedaddress]{revtex4}

\usepackage{graphicx}  
\usepackage{dcolumn}   
\usepackage{bm}        
\usepackage{amssymb}   
\usepackage{setspace}

\def\lsim{\mathrel{\rlap{\lower4pt\hbox{\hskip1pt$\sim$}}\raise1pt\hbox{$<$}}}
\def\gsim{\mathrel{\rlap{\lower4pt\hbox{\hskip1pt$\sim$}}\raise1pt\hbox{$>$}}}
\def\MET{{\mbox{$E\kern-0.57em\raise0.19ex\hbox{/}_{T}$}}}
\def\METwithSpace{{\mbox{$E\kern-0.57em\raise0.19ex\hbox{/}_{T}~$}}}

\begin{document}

\widetext


\input author_list.tex       

\date{June 3, 2010}

\title{Measurement of the $WZ\rightarrow \ell\nu\ell\ell$ cross section
and limits on anomalous triple gauge couplings in $p\bar{p}$ collisions at $\sqrt{s}$~=~1.96~TeV}

\begin{abstract}

We present a new measurement of the $WZ\rightarrow \ell\nu\ell\ell$ ($\ell = e,\mu$)
cross section and limits on anomalous triple gauge couplings. Using 4.1~fb$^{-1}$ 
of integrated luminosity of $p\bar{p}$ collisions at $\sqrt{s} = 1.96$~TeV, 
we observe 34 $WZ$ candidate events with an estimated background of 
$6.0 \pm 0.4$ events. We measure the $WZ$ production cross 
section to be 
$3.90^{+1.06}_{-0.90}$~pb,
in good agreement with the standard model prediction.
We find no evidence for anomalous $WWZ$ couplings and set 
95\%~C.L. limits on the coupling parameters, $-0.075 < \lambda_{Z} < 0.093$ 
and $-0.027 < \Delta\kappa_{Z} < 0.080$, in the HISZ parameterization for a
$\Lambda = 2$~TeV form factor scale. These are the best limits to date obtained 
from the direct measurement of the $WWZ$ vertex.

\end{abstract}

\maketitle


The standard model (SM) of particle physics has been extensively tested in 
the past three decades and is found to be in excellent agreement with 
experimental observations. It is widely assumed, however, that the SM is 
only a low energy approximation of a more general theory.
Therefore, any significant deviation from the SM predictions yields information on
the nature of a more fundamental theory. Production of $WZ$ pairs is the 
least studied diboson process within the SM, as it is a charged final state 
and can only be produced at hadron colliders. 
A detailed study of this process probes the electroweak sector of the SM.
In addition, searches for new phenomena in the production
of heavy gauge boson pairs are interesting, as
many extensions of the SM predict~\cite{ssm, extradim, litHig, LSTC} additional
heavy gauge bosons that can decay into a $WZ$ boson pairs.

In the SM, $WZ$ boson pairs are produced at leading order (LO) via 
$t$-,~$u$-, and $s$-channels.
These channels interfere and maintain unitarity at high energies. 
In the case of the $t$- and $u$-channels, the 
$W$ and $Z$ bosons are radiated from initial state quarks, while 
the $s$-channel production occurs via the $WWZ$ triple gauge boson vertex, which 
is a consequence of the non-Abelian nature of the SM. 
There are 14 free parameters describing the generalized Lagrangian for the
$WWV$ interaction~\cite{Th-Hagiw1,Th-Hagiw2}, where $V$ is either a $Z$ boson
or a photon. Assuming gauge invariance and 
conservation of the $C$, $P$, and $CP$ symmetries, only six remain.
Their notation and SM values are
$\lambda_{V} = 0$, $\kappa_{V} = g_{1}^{V} = 1$ for the $WWV$ vertex,
while the deviations from the SM values are noted as 
$\Delta\kappa_{V}$, $\Delta g_{1}^{V}$,~and~$\lambda_{V}$.
The $U(1)$ electromagnetic gauge invariance implies $\Delta g^{\gamma}_1 = 0$.
In this Letter, we describe the $WWZ$ vertex in three-dimensional (3D) phase space
of coupling parameters, $\Delta\kappa_{Z}$, $\Delta g_{1}^{Z}$,~and~$\lambda_{Z}$.
We also consider the HISZ parameterization~\cite{Coup-HISZ} that implies
$\Delta\kappa_Z = \Delta g^{Z}_1({\rm cos}^2\theta_{W}-{\rm sin}^2\theta_{W})$.
Thus, the $WWZ$ vertex can be described by
$\Delta\kappa_Z$ and $\lambda_Z$ only.

If the coupling parameters have non-SM values, new physics is required to 
prevent gauge boson production from violating unitarity at high energies.
The high energy behavior is controlled by introducing a dipole form factor scale,
$\Lambda$, in the description of the couplings,
$\alpha(\hat{s}) \to \alpha_0/(1+\hat{s}/\Lambda^2)^2$,
where $\hat{s}$ is the square of the partonic center-of-mass energy and $\alpha_0$ is the coupling
value in the low energy approximation.

The $WZ$ production cross section was previously measured to be 
$\sigma(p\bar{p}\to WZ) = 5.0^{+1.8}_{-1.6}$~pb~\cite{CDF-CrossSec} and  
$\sigma(p\bar{p}\to WZ) = 2.7^{+1.7}_{-1.3}$~pb~\cite{D0RunIIaWZ}, by the CDF and 
D0 collaborations, respectively, using $\sim $1~fb$^{-1}$ of integrated  
luminosity.
Combined limits on the gauge couplings from the CERN LEP 
collider were obtained~\cite{LEP} by the indirect measurement 
of the $WWZ$ coupling in the $e^{+}e^{-}\to W^{+}W^{-}$ process. The only direct measurement  
of $WWZ$ couplings was performed at the Tevatron. 
Using 1~fb$^{-1}$ of integrated luminosity, 95\% C.L. limits 
on anomalous $WWZ$ couplings were derived~\cite{D0RunIIaWZ} by the D0 experiment:
$-0.17 < \lambda_{Z} < 0.21$,  $-0.14 < \Delta g^{Z}_{1} < 0.34$ for the HISZ relation
and $-0.12 < \Delta\kappa_{Z} = \Delta g_{1}^{Z} < 0.29$, using $\Lambda = 2$~TeV.
The CDF experiment used data equivalent to 350~pb$^{-1}$ of integrated luminosity 
that resulted in 95\% C.L. limits on anomalous $WWZ$ couplings~\cite{CDF-coup}:
$-0.28 < \lambda_{Z} < 0.28$ and $-0.50 < \Delta\kappa_{Z} < 0.43$ assuming
equal coupling relation between $WWZ$ and $WW\gamma$ couplings and 
$\Lambda = 1.5$~TeV.
 
In this Letter, we present a new measurement of the $WZ$ production cross section
and set 95\%~C.L. limits on the deviation from the SM predictions of triple gauge couplings 
($\lambda_{Z} $, $\Delta\kappa_{Z}$, $\Delta g_{1}^{Z} $) using data equivalent to 
4.1~fb$^{-1}$ of integrated luminosity of $p\bar{p}$ collisions at $\sqrt{s} = 1.96$~TeV
at the Tevatron collected by the D0 detector. 
This supersedes the previous D0 measurement.
We consider only the leptonic decays of the $W$ and $Z$ bosons into final states 
with electrons, muons, and with missing transverse energy (\MET)~\cite{delta_r} 
due to the neutrino from the $W$ boson decay. 

The detailed description of the D0 detector can be found elsewhere~\cite{run2det}, while here
we present a brief overview of the main sub-systems of the detector. The inner most part 
is a central tracking system surrounded  by a 2~T superconducting solenoidal magnet. 
The two components of the central tracking system, a silicon microstrip tracker and a central 
fiber tracker, are used to reconstruct interaction vertexes and provide the measurement 
of the momentum of charged particles.
The tracking system and a magnet are followed by the calorimetry system that consists of
central (CC) and endcap (EC) electromagnetic and hadronic uranium-liquid argon
sampling calorimeters, and an intercryostat detector (ICD).
A central calorimeter and two endcap calorimeters cover the
pseudorapidity ranges $|\eta| < 1.1$ and $1.5 < |\eta| < 4.2$, respectively, while
the ICD provides coverage for $1.1 < |\eta| < 1.4$.
The calorimeter measures energy of hadrons, electrons, and photons.
Outside of the D0 calorimeter lies a muon system which consists of layers of drift 
tubes and scintillation counters and a 1.8~T toroidal magnet.

An electron candidate is identified as a cluster of energy in the CC, EC, or ICD that is matched to a track reconstructed in the D0 central tracker. Due to different coverage of the tracker, we select EC electrons within $1.5 < |\eta| < 2.5$ and CC electrons within $|\eta|< 1.1$. The cluster in the CC or EC must be isolated and have a shower shape consistent with that of an electron. In the intercryostat region (ICR), $1.1 < |\eta| < 1.5$, we cluster energy found in the CC, ICD, or EC detectors. These ICR electrons are required to pass a neural network discriminant that uses the cluster's shower shape
and associated track information. A muon candidate is reconstructed as segments within the muon system that are matched to a track reconstructed in the central tracker. 
The muon candidate track must be isolated from activity in the tracker and the calorimeter.

The Monte Carlo (MC) samples of $WZ$ signal and $ZZ$ background are produced 
using the {\sc pythia}~\cite{pythia} generator. The production of the $W$ and $Z$ bosons
in association with jets ($W + $jets, $Z + $jets), collectively referred to as $V + $jets,
and $t\bar{t}$ processes are generated using {\sc alpgen}~\cite{alpgen}
interfaced with {\sc pythia} for showering and hadronization.
All MC samples are passed through the {\sc geant}~\cite{geant} simulation of the D0 detector. 
The simulated samples are further corrected to describe the luminosity dependence of 
the trigger and reconstruction efficiencies in data, as well as 
the beam spot position. All MC samples are normalized 
to the luminosity in data using next-to-leading order (NLO) calculations of the 
cross sections and 
are subject to the same selection criteria as that applied to data.

We consider four independent decay signatures: $eee+\MET$,
$ee\mu+\MET$, $\mu\mu e+\MET$, and $\mu\mu\mu+\MET$.
Electron reconstructed in the ICR
must be selected as one of the electrons from the 
$Z$ boson decay. We require the events to have at least three
lepton candidates with $p_T > 15$~GeV that originate
from the same vertex and separated from each other by at least 
$\Delta R = \sqrt{(\Delta\phi)^2+ (\Delta\eta)^2} > 0.5$.
The event must also have a significant $\MET$ to account for the unobserved
neutrino. We require $\MET$ to be above 20 GeV.
Events are selected using triggers based on
electrons and muons. Since there are multiple high $p_T$ leptons from the decay of
the heavy gauge bosons the trigger efficiency is measured to be 
$98\% \pm 2\%$ for all signatures.

In the $WZ$ candidate selection, we first identify the leptons from the $Z$ boson 
decay. 
We consider all pairs of electrons or muons, additionally requiring opposite electrical charge in the cases of muon pairs or electron pairs including an ICR electron.
The pair that has an invariant mass closest to and consistent with the 
$Z$ boson nominal mass is selected as coming from the $Z$ boson decay.
If such pair is not found the event is rejected.
The lepton from the $W$ boson decay is selected as the one with
the highest transverse momentum from the remaining unassigned muons and 
CC or EC electrons in the event.
This assignment is studied in the simulation and found to be 100\% correct for
$ee\mu$ and $\mu\mu e$ channels.
It is found to be correct in about 92\% and 89\% of cases for $eee$ and $\mu\mu\mu$ signatures, respectively. 
The effects of misassignment on the product of acceptance and efficiency
of the selection criteria, $\cal{A}\times\epsilon$, are estimated in
the signal simulation. Values of $\cal{A}\times\epsilon$ measured using the
assignment method described above differ from those obtained using
MC generator-level information by less than one per cent. Therefore, the
systematic uncertainty on $\cal{A}\times\epsilon$ due to the misassignment is 
neglected in this analysis.

In order to reduce the background contamination,
the thresholds in the selection criteria are further optimized for each $WZ$ decay mode
by maximizing $S/\sqrt{S+B}$.
Here, $S$ is the expected number of $WZ$ signal events and $B$ is the total 
number of background events. 
The simulation is used to estimate $S$ as well as to measure 
$\cal{A}\times\epsilon$ for
each decay signature. The kinematic selection criteria are applied 
to measure the acceptance in simulations, while the 
lepton identification efficiencies are measured in data.
The results are summarized in Table~\ref{Axeff}.

\begin{table}
\begin{center}
\begin{tabular}{lc}\hline\hline
Channel             & $\cal{A}\times\epsilon$ (\%) \\ \hline
$eee$                 &  $1.35 \pm 0.15$          \\ 
$ee\mu$             & $1.57  \pm 0.12$          \\ 
$\mu \mu e$       & $1.07  \pm 0.11$          \\ 
$\mu \mu \mu$  &  $1.34  \pm 0.13$         \\ \hline\hline
\end{tabular}
\caption{Acceptance multiplied by efficiency, $\cal{A}\times\epsilon$, of the full selection criteria 
for each decay signature. $\cal{A}\times\epsilon$ values are calculated with respect to the fully 
leptonic $WZ$ decay simulation. The uncertainties are both statistical and systematic.}
\label{Axeff}
\end{center}
\end{table}

The major background is from processes with a 
$Z$ boson and an additional object misidentified as the lepton from the 
$W$ boson decay. Such processes are $Z + $jets, $ZZ$, and $Z \gamma$.   
A small background contribution is expected from processes without 
$Z$ boson, such as $W + $jets and $t\bar{t}$ processes.

The $ZZ$ and $t\bar{t}$ backgrounds are estimated from the simulation,
while the $V + $jets, with $V$ being either a $Z$ or $W$ bosons, and $Z\gamma$ 
backgrounds are estimated using data-driven methods.

One or more jets in the $V + $jets process can be misidentified as a lepton from the 
$W$ or $Z$ boson decays. To estimate this contribution, we define a {\it false}
lepton category for electrons and muons.
A {\it false} electron is required to have most of its energy deposited in the 
electromagnetic calorimeter and satisfy electron calorimeter isolation criteria, while 
having a shower shape inconsistent with that of an electron. 
A muon candidate is categorized as {\it false} if it fails the isolation criteria. 
These requirements ensure that the 
{\it false} lepton is either a misidentified jet or a lepton from the semi-leptonic decay
of heavy flavor quarks. Using a multijet data sample, we measure the 
ratio of misidentified leptons passing two different selection
criteria, {\it false} lepton and signal lepton, as a function of $p_T$ and $\eta$ for
electrons and muons, respectively.
We then select a sample of $Z$ boson decays with an additional 
{\it false} lepton candidate for each final state signature. The contribution from
the $V + $jets background is estimated 
by scaling the number of events in this sample
by the corresponding $p_T$- or $\eta$-dependent misidentification ratio.

Initial or final state radiation in $Z\gamma$ events can mimic the signal process if
the photon either converts into $e^{+}e^{-}$ pair or when a central track is wrongly 
matched to a photon. As a result, the $Z\gamma$ process is a background to 
two out of the four final state signatures with $W\to e\nu$ decays.  
To estimate the contribution from this background, we measure the rate at 
which a photon is misidentified as an electron. This is estimated using a
data sample of $Z\to\mu\mu$ events with a final state radiation photon,
since it offers an almost background-free source of photons due to the
invariant mass, $M(\mu\mu\gamma)$, constraint to the $Z$ boson mass. 
The muon decay of the $Z$ boson is chosen to avoid an ambiguity when
assigning the electromagnetic shower to the final state photon candidate.
The misidentification rate is measured as a function of the $p_T$ of the
electromagnetic shower.
The $Z\gamma$ contribution is estimated by multiplying the $p_{T}$-dependent 
misidentification rate by the photon $p_T$ 
distribution in the $Z\gamma$ NLO MC simulation~\cite{Baur}.

\begin{figure}
\begin{center}
\includegraphics[scale=0.4]{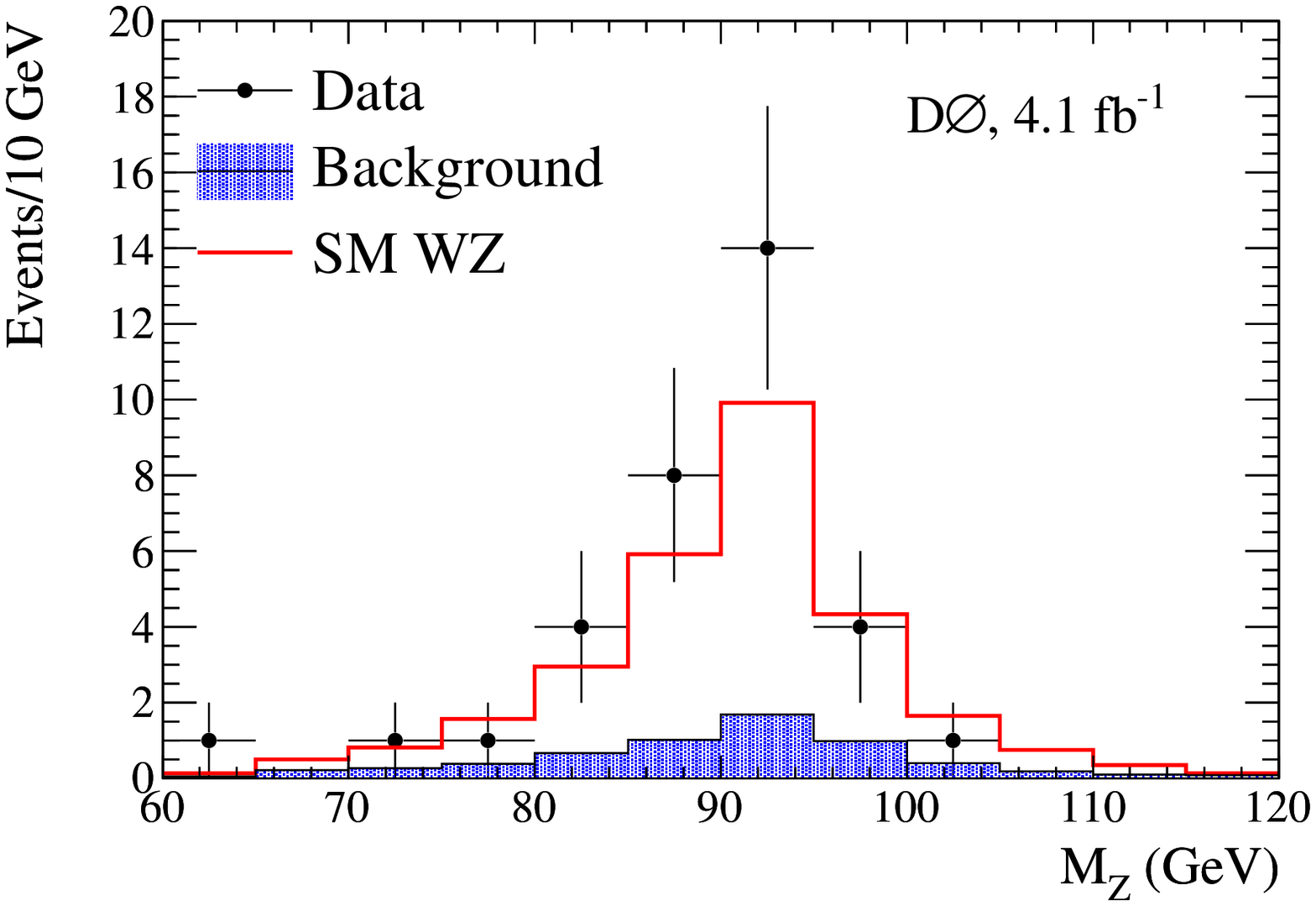}
\caption{(Color online) Invariant mass distribution of selected $Z$ candidates in data (black points), 
   with $WZ$ signal (open histogram) and total background (dark histogram) overlaid.}
\label{fig:zmass}
\end{center}
\end{figure}

\begin{figure}
\begin{center}
\includegraphics[scale=0.4]{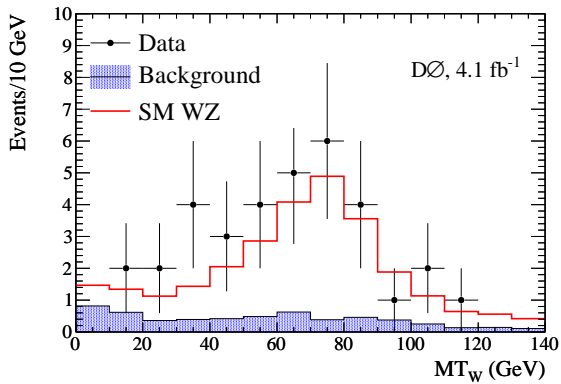}
\caption{(Color online) Transverse mass distribution of selected $W$ candidates in data (black points), 
   with $WZ$ signal (open histogram) and total background (dark histogram) overlaid.}
\label{fig:wtrmass}
\end{center}
\end{figure}

The selection yields 34 $WZ$ candidate events with an estimated 
$23.3 \pm 1.5$ signal, and $6.0 \pm 0.6$ 
background events.
The number of observed candidate events as well as the expected numbers of signal 
and background events for each signature are 
summarized in Table~\ref{tab:EvYields}. 
The distribution of the invariant mass of the $Z$ boson candidates is given in Fig.~\ref{fig:zmass}.
The transverse mass of the $W$ boson candidate is calculated as follows
\begin{equation}
MT_{W} = \sqrt{2E_{T}^{\ell}\MET(1-{\rm cos}(\phi^{\ell}-\phi^{\MET}))},
\end{equation}
where $E_{T}^{\ell}$ and $\phi^{\ell}$ are transverse energy and 
azimuthal angle, respectively, of the electron or muon selected as the $W$ boson 
decay product and $\phi^{\MET}$ is the azimuthal angle of the missing transverse momentum.
The distribution of the $W$ boson candidates is given in Fig.~\ref{fig:wtrmass}.

\begin{table*}[ht] 
\begin{center} 
\begin{tabular}{lcccc} \hline \hline
Source            & $eee$                    & $ee\mu$                  & $e\mu\mu$                & $\mu\mu\mu$     \\ \hline
$ZZ$                & $0.39 \pm 0.07$ & $1.48 \pm 0.20$ & $0.40 \pm 0.07$ & $1.26 \pm 0.23$ \\
$V + $jets       & $0.63 \pm 0.17$ &  $0.56 \pm 0.24$ & $0.03 \pm 0.01$ & $0.17 \pm 0.05$ \\
$Z\gamma$    & $0.28 \pm 0.08$ & $< 0.001$             & $0.66 \pm 0.34$ & $< 0.001$          \\
$t\bar{t}$         & $0.03 \pm 0.01$ & $0.05 \pm 0.01$  & $0.04 \pm 0.01$ & $0.03 \pm 0.01$ \\ \hline
Total bkg.        & $1.33 \pm 0.21$ & $2.11 \pm 0.31$ & $1.13 \pm 0.35$ & $1.46 \pm 0.24$  \\
$WZ$ signal   & $5.9 \pm 0.8$   & $6.9 \pm 0.8$   & $4.7 \pm 0.6$   & $5.8 \pm 0.8$            \\
Observed            & 9                        & 11                        & 9                        & 5                       \\ \hline \hline 
\end{tabular}
\end{center} 
\caption{Number of observed events, expected number of signal events, 
        and expected number of background events for each final state signature 
        with total (statistical and systematic) uncertainties.}  
\label{tab:EvYields}
\end{table*}

Several sources of systematic uncertainty are considered.
The systematic uncertainties on the lepton identification efficiencies are 
5\%, 4\%, and 6\% for CC/EC electrons, muons, and ICR electrons, respectively.
The systematic uncertainty assigned to the PDF choice is 5\%.
A systematic uncertainty of 5\% is assigned on $\cal{A}\times\epsilon$ due to 
modeling of the kinematics of the $WZ$ system.
In addition, we assign 
7\%~\cite{ttbar-cross-sec} and 10\%~\cite{zz-cross-sec} 
systematic uncertainty to the estimated $t\bar{t}$ and $ZZ$ backgrounds, respectively,  
due to the uncertainty on their theoretical cross sections.
The major sources of systematic uncertainty on the estimated $V + $jets
contribution are the \MET~requirement and the statistics in the multijet sample used to measure 
the lepton misidentification ratios. These effects are estimated independently for each signature 
and found to be between 20-30\%. 
The systematic uncertainty on the $Z\gamma$ background is estimated
to be 40\% and 58\% for the $eee$ and $\mu\mu e$ channels, respectively. 

A likelihood method~\cite{lhood} is used to combine the four measurements, taking
into account the correlations among the systematic uncertainties on the expected
signal and the estimated background contributions.
The cross section is $\sigma (WZ)~=~3.90^{+1.01}_{-0.85}~({\rm stat + syst})~\pm 0.31~({\rm lumi})$~pb.
The uncertainties are dominated by the statistics of the number of observed candidates.
The luminosity uncertainty includes 6.1\% relative uncertainty~\cite{lumi} due to the
luminosity measurement and the normalization uncertainty of the background contributions
estimated from MC simulation.

\begin{figure}[htbp]
 \begin{center}  	
	{\includegraphics[width=0.5\textwidth]{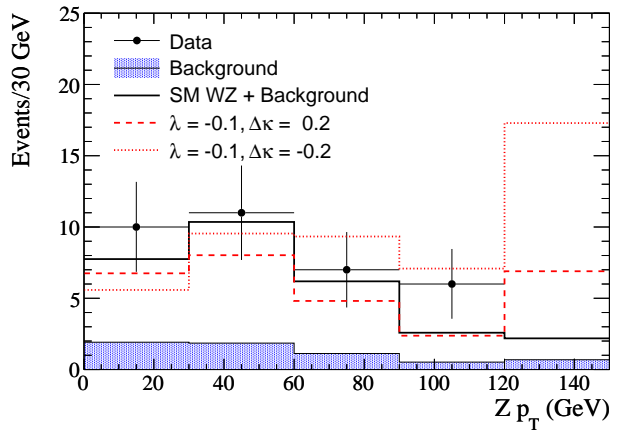}}
    \caption{(Color online) The $Z$ boson $p_T$ spectrum from data (points), total background (dark histogram), 
	the SM $WZ$ single + total background (open histogram), and two anomalous coupling models (dashed and dotted histograms).
	The last bin includes overflows.}
    \label{fig:zptspec}  
  \end{center}  
\end{figure}

The presence of anomalous $WWZ$ couplings would lead to both an increase
in the cross section and a change in the $p_T$ spectrum of the $W$ and $Z$ bosons.
We use the $Z$ boson $p_T$ distribution to set limits on the
coupling parameters using a form factor scale $\Lambda = 2$~TeV.
The $Z$ boson $p_T$ spectra from data, the SM, and two anomalous coupling predictions are 
shown in Fig.~\ref{fig:zptspec}.
The difference is most pronounced in the last bin, which includes also the events
above 150~GeV.

\begin{figure}[htbp]
 \begin{center}  	
	{\includegraphics[width=0.4\textwidth]{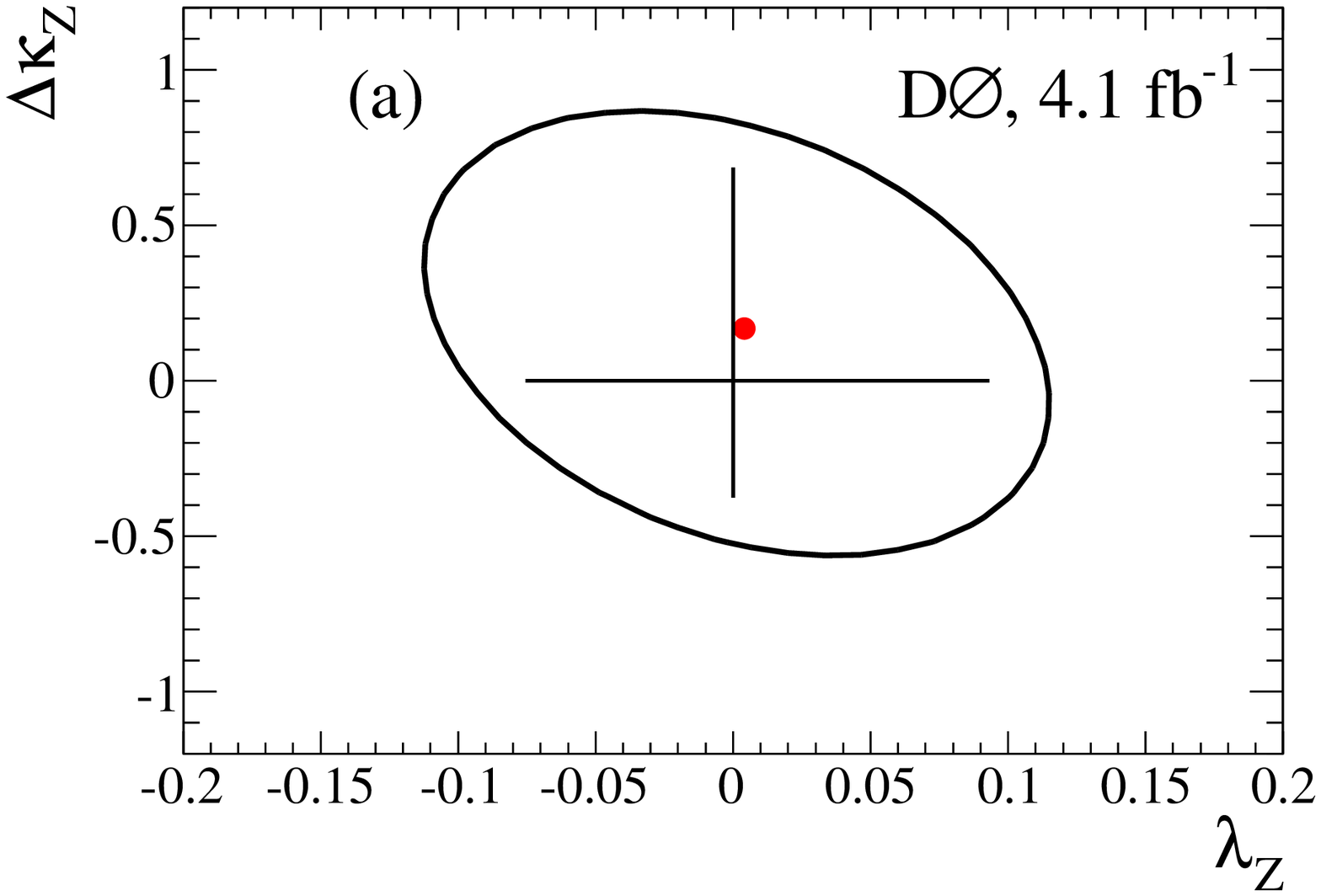}}	
	{\includegraphics[width=0.4\textwidth]{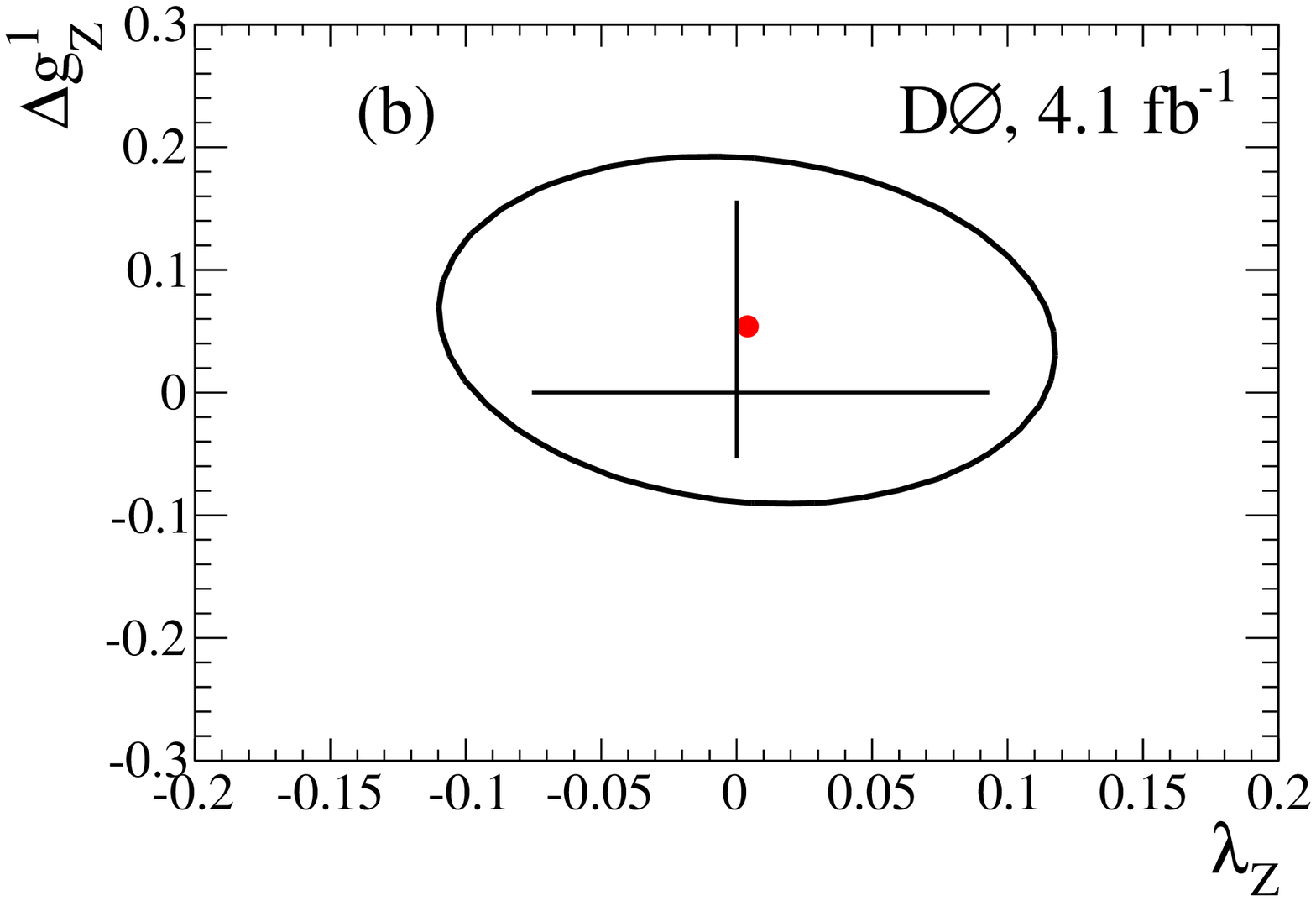}}	
	{\includegraphics[width=0.4\textwidth]{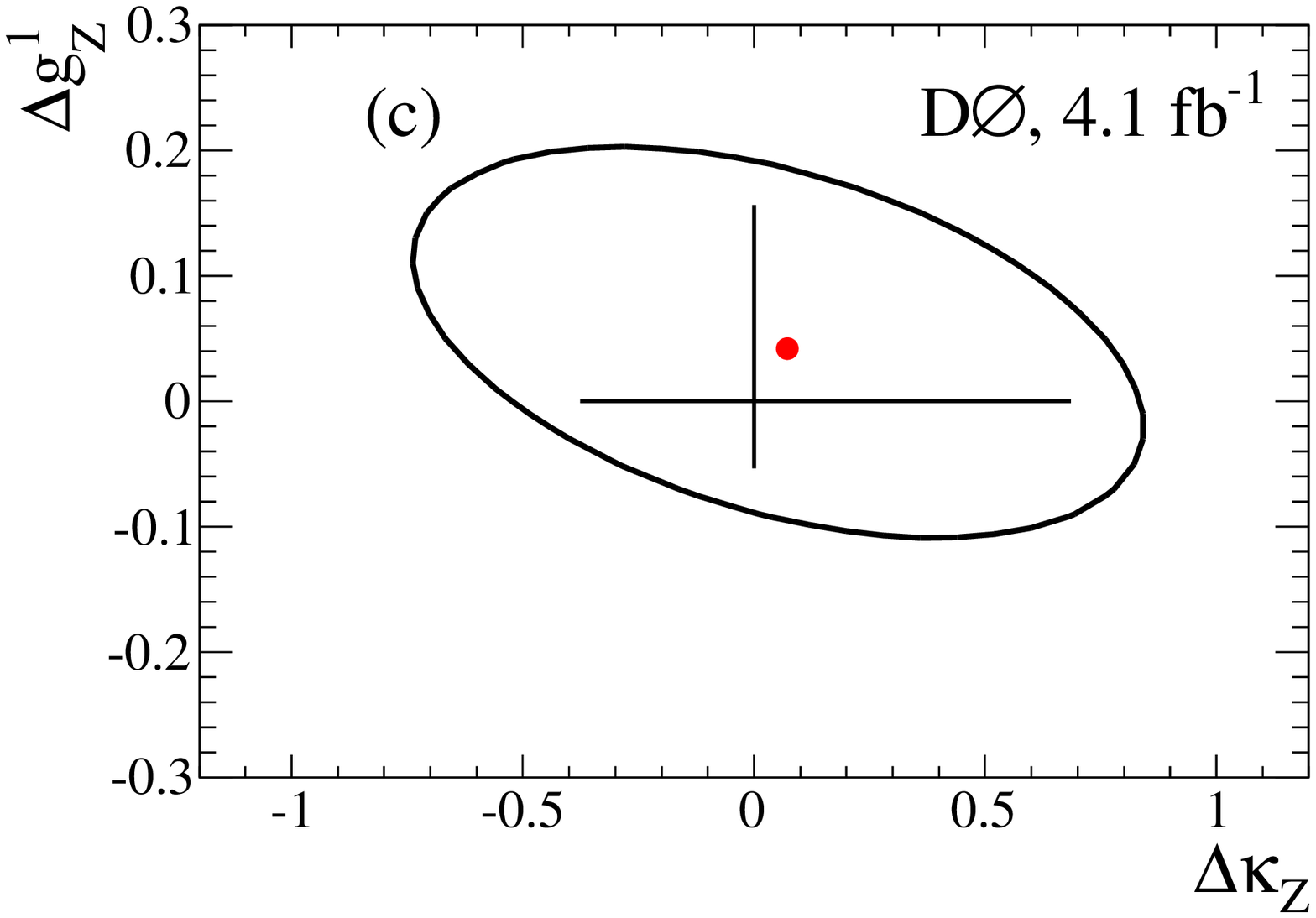}}
    \caption{(Color online) Two-dimensional 95\% C.L limit contours on
    	$(\Delta\kappa_{Z},\lambda_{Z})$ (a), $(\Delta g_{1}^{Z},\lambda_{Z})$ (b), and 
$(\Delta g_{1}^{Z},\Delta\kappa_{Z})$ (c).
		The point corresponds to the minimum of the likelihood surface. 
		The vertical and horizontal lines represent the one-dimensional limits calculated separately.
		A form factor scale of 2~TeV is used.}
    \label{fig:limits_lep}  
  \end{center}  
\end{figure}

\begin{figure}[htbp]
 \begin{center}  	
	{\includegraphics[width=0.4\textwidth]{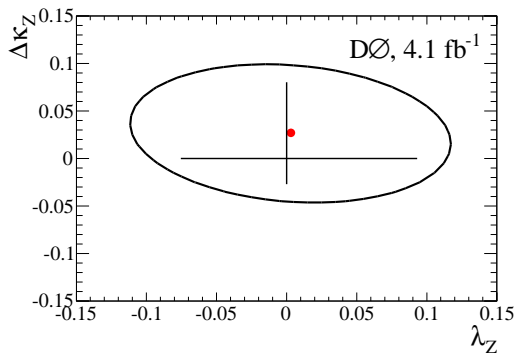}}	
    \caption{(Color online) Two-dimensional 95\% C.L limit contours for the HISZ parameterization. The point corresponds to the minimum of the likelihood surface. The vertical and horizontal lines represent the separately calculated one-dimensional limits.}
    \label{fig:limits_hisz}  
  \end{center}  
\end{figure}

A three-dimensional grid of values of anomalous couplings
$\Delta\kappa_{Z}$, $\Delta g_{1}^{Z}$,~and~$\lambda_{Z}$
is produced. For each point of the grid we generate $WZ$ production
using {\sc mcfm}~\cite{zz-cross-sec} and obtain normalized to luminosity
$p_T$ spectrum of the $Z$ boson. This spectrum combined with that
from the estimated background is compared with the measured $Z$ boson 
$p_T$ spectrum in data. The likelihood of the match is calculated 
with the assumption of Poisson statistics for the signal and Gaussian uncertainties for 
the background. The two-dimensional 95\%~C.L. limit contours in three 
planes, $(\Delta\kappa_{Z},\lambda_{Z})$, $(\Delta g_{1}^{Z},\lambda_{Z})$, and
$(\Delta g_{1}^{Z},\Delta\kappa_{Z})$, are shown in Fig.~\ref{fig:limits_lep}. In each case the
third coupling is restricted to the SM value.
For the HISZ parameterization the results are presented as 
limits on two coupling parameters: $\Delta\kappa_{Z}$ and $\lambda_{Z}$.
The corresponding two-dimensional 95\% C.L. limit contour
is shown on Fig.~\ref{fig:limits_hisz}. 
The one-dimensional limits on the coupling parameters obtained without any coupling relation 
and with HISZ parameterization are summarized in Table~\ref{tab:OneDLim}. 

\begin{table}[ht] 
\begin{center} 
\begin{tabular}{lc} \hline \hline
Coupling relation & 95\%~C.L. Limit\\ \hline
$\Delta g_{1}^{Z}=\Delta\kappa_{Z}=0$     & $ -0.075 < \lambda_{Z} < 0.093$ \\
$\lambda_{Z}=\Delta\kappa_{Z}=0$           & $ -0.053 <  \Delta g_{1}^{Z} < 0.156$ \\ 
$\lambda_{Z}=\Delta g_{1}^{Z}=0$            & $ -0.376 <  \Delta\kappa_{Z} < 0.686$ \\ \hline
$\Delta\kappa_{Z}=0$ (HISZ)                    & $ -0.075 < \lambda_{Z} < 0.093$ \\
$\lambda_{Z}=0$ (HISZ)                           & $ -0.027 < \Delta\kappa_{Z} < 0.080$\\ \hline \hline
\end{tabular}
\end{center} 
\caption{One-dimensional 95\%~C.L. limits on anomalous coupling parameters obtained from
varying one of the couplings while fixing the remaining couplings to the SM values (top three results).
The last two results correspond to one-dimensional 95\%~C.L. limits on anomalous coupling
parameters for the HISZ parameterization. A form factor scale of $\Lambda =2$~TeV is used.}
\label{tab:OneDLim}
\end{table}

In summary, we have presented a measurement of the $WZ$ production cross section using
4.1~fb$^{-1}$ of integrated luminosity of D0 data.
We observe 34 events with $23.3 \pm 1.5$ expected signal events and $6.0 \pm 0.6$
estimated background events. We measure the $WZ$ cross section to be 
$3.90^{+1.06}_{-0.90}$~pb, which is in
agreement with the SM NLO prediction of $3.25 \pm 0.19$~pb~\cite{zz-cross-sec}. This is the most
precise measurement to date of the $WZ$ cross section. We find no evidence
for anomalous $WWZ$ couplings and set 95\% C.L. limits of
$-0.075 < \lambda_{Z} < 0.093$ and $-0.027 < \Delta\kappa_{Z} < 0.080$ for the
HISZ parametrization using $\Lambda = 2$~TeV. 
These are the most stringent limits on $WWZ$ couplings
obtained from the study of direct $WZ$ production. 


\input acknowledgement.tex
\end{document}

%% file: author_list.tex
%
\affiliation{Universidad de Buenos Aires, Buenos Aires, Argentina}
\affiliation{LAFEX, Centro Brasileiro de Pesquisas F{\'\i}sicas, Rio de Janeiro, Brazil}
\affiliation{Universidade do Estado do Rio de Janeiro, Rio de Janeiro, Brazil}
\affiliation{Universidade Federal do ABC, Santo Andr\'e, Brazil}
\affiliation{Instituto de F\'{\i}sica Te\'orica, Universidade Estadual Paulista, S\~ao Paulo, Brazil}
\affiliation{Simon Fraser University, Vancouver, British Columbia, and York University, Toronto, Ontario, Canada}
\affiliation{University of Science and Technology of China, Hefei, People's Republic of China}
\affiliation{Universidad de los Andes, Bogot\'{a}, Colombia}
\affiliation{Charles University, Faculty of Mathematics and Physics, Center for Particle Physics, Prague, Czech Republic}
\affiliation{Czech Technical University in Prague, Prague, Czech Republic}
\affiliation{Center for Particle Physics, Institute of Physics, Academy of Sciences of the Czech Republic, Prague, Czech Republic}
\affiliation{Universidad San Francisco de Quito, Quito, Ecuador}
\affiliation{LPC, Universit\'e Blaise Pascal, CNRS/IN2P3, Clermont, France}
\affiliation{LPSC, Universit\'e Joseph Fourier Grenoble 1, CNRS/IN2P3, Institut National Polytechnique de Grenoble, Grenoble, France}
\affiliation{CPPM, Aix-Marseille Universit\'e, CNRS/IN2P3, Marseille, France}
\affiliation{LAL, Universit\'e Paris-Sud, CNRS/IN2P3, Orsay, France}
\affiliation{LPNHE, Universit\'es Paris VI and VII, CNRS/IN2P3, Paris, France}
\affiliation{CEA, Irfu, SPP, Saclay, France}
\affiliation{IPHC, Universit\'e de Strasbourg, CNRS/IN2P3, Strasbourg, France}
\affiliation{IPNL, Universit\'e Lyon 1, CNRS/IN2P3, Villeurbanne, France and Universit\'e de Lyon, Lyon, France}
\affiliation{III. Physikalisches Institut A, RWTH Aachen University, Aachen, Germany}
\affiliation{Physikalisches Institut, Universit{\"a}t Freiburg, Freiburg, Germany}
\affiliation{II. Physikalisches Institut, Georg-August-Universit{\"a}t G\"ottingen, G\"ottingen, Germany}
\affiliation{Institut f{\"u}r Physik, Universit{\"a}t Mainz, Mainz, Germany}
\affiliation{Ludwig-Maximilians-Universit{\"a}t M{\"u}nchen, M{\"u}nchen, Germany}
\affiliation{Fachbereich Physik, Bergische  Universit{\"a}t Wuppertal, Wuppertal, Germany}
\affiliation{Panjab University, Chandigarh, India}
\affiliation{Delhi University, Delhi, India}
\affiliation{Tata Institute of Fundamental Research, Mumbai, India}
\affiliation{University College Dublin, Dublin, Ireland}
\affiliation{Korea Detector Laboratory, Korea University, Seoul, Korea}
\affiliation{CINVESTAV, Mexico City, Mexico}
\affiliation{FOM-Institute NIKHEF and University of Amsterdam/NIKHEF, Amsterdam, The Netherlands}
\affiliation{Radboud University Nijmegen/NIKHEF, Nijmegen, The Netherlands}
\affiliation{Joint Institute for Nuclear Research, Dubna, Russia}
\affiliation{Institute for Theoretical and Experimental Physics, Moscow, Russia}
\affiliation{Moscow State University, Moscow, Russia}
\affiliation{Institute for High Energy Physics, Protvino, Russia}
\affiliation{Petersburg Nuclear Physics Institute, St. Petersburg, Russia}
\affiliation{Stockholm University, Stockholm and Uppsala University, Uppsala, Sweden }
\affiliation{Lancaster University, Lancaster LA1 4YB, United Kingdom}
\affiliation{Imperial College London, London SW7 2AZ, United Kingdom}
\affiliation{The University of Manchester, Manchester M13 9PL, United Kingdom}
\affiliation{University of Arizona, Tucson, Arizona 85721, USA}
\affiliation{University of California Riverside, Riverside, California 92521, USA}
\affiliation{Florida State University, Tallahassee, Florida 32306, USA}
\affiliation{Fermi National Accelerator Laboratory, Batavia, Illinois 60510, USA}
\affiliation{University of Illinois at Chicago, Chicago, Illinois 60607, USA}
\affiliation{Northern Illinois University, DeKalb, Illinois 60115, USA}
\affiliation{Northwestern University, Evanston, Illinois 60208, USA}
\affiliation{Indiana University, Bloomington, Indiana 47405, USA}
\affiliation{Purdue University Calumet, Hammond, Indiana 46323, USA}
\affiliation{University of Notre Dame, Notre Dame, Indiana 46556, USA}
\affiliation{Iowa State University, Ames, Iowa 50011, USA}
\affiliation{University of Kansas, Lawrence, Kansas 66045, USA}
\affiliation{Kansas State University, Manhattan, Kansas 66506, USA}
\affiliation{Louisiana Tech University, Ruston, Louisiana 71272, USA}
\affiliation{University of Maryland, College Park, Maryland 20742, USA}
\affiliation{Boston University, Boston, Massachusetts 02215, USA}
\affiliation{Northeastern University, Boston, Massachusetts 02115, USA}
\affiliation{University of Michigan, Ann Arbor, Michigan 48109, USA}
\affiliation{Michigan State University, East Lansing, Michigan 48824, USA}
\affiliation{University of Mississippi, University, Mississippi 38677, USA}
\affiliation{University of Nebraska, Lincoln, Nebraska 68588, USA}
\affiliation{Rutgers University, Piscataway, New Jersey 08855, USA}
\affiliation{Princeton University, Princeton, New Jersey 08544, USA}
\affiliation{State University of New York, Buffalo, New York 14260, USA}
\affiliation{Columbia University, New York, New York 10027, USA}
\affiliation{University of Rochester, Rochester, New York 14627, USA}
\affiliation{State University of New York, Stony Brook, New York 11794, USA}
\affiliation{Brookhaven National Laboratory, Upton, New York 11973, USA}
\affiliation{Langston University, Langston, Oklahoma 73050, USA}
\affiliation{University of Oklahoma, Norman, Oklahoma 73019, USA}
\affiliation{Oklahoma State University, Stillwater, Oklahoma 74078, USA}
\affiliation{Brown University, Providence, Rhode Island 02912, USA}
\affiliation{University of Texas, Arlington, Texas 76019, USA}
\affiliation{Southern Methodist University, Dallas, Texas 75275, USA}
\affiliation{Rice University, Houston, Texas 77005, USA}
\affiliation{University of Virginia, Charlottesville, Virginia 22901, USA}
\affiliation{University of Washington, Seattle, Washington 98195, USA}
\author{V.M.~Abazov} \affiliation{Joint Institute for Nuclear Research, Dubna, Russia}
\author{B.~Abbott} \affiliation{University of Oklahoma, Norman, Oklahoma 73019, USA}
\author{M.~Abolins} \affiliation{Michigan State University, East Lansing, Michigan 48824, USA}
\author{B.S.~Acharya} \affiliation{Tata Institute of Fundamental Research, Mumbai, India}
\author{M.~Adams} \affiliation{University of Illinois at Chicago, Chicago, Illinois 60607, USA}
\author{T.~Adams} \affiliation{Florida State University, Tallahassee, Florida 32306, USA}
\author{G.D.~Alexeev} \affiliation{Joint Institute for Nuclear Research, Dubna, Russia}
\author{G.~Alkhazov} \affiliation{Petersburg Nuclear Physics Institute, St. Petersburg, Russia}
\author{A.~Alton$^{a}$} \affiliation{University of Michigan, Ann Arbor, Michigan 48109, USA}
\author{G.~Alverson} \affiliation{Northeastern University, Boston, Massachusetts 02115, USA}
% ...
%
%
\collaboration{The D0 Collaboration\footnote{with visitors from
$^{a}$Augustana College, Sioux Falls, SD, USA,
$^{b}$The University of Liverpool, Liverpool, UK,
$^{c}$SLAC, Menlo Park, CA, USA,
$^{d}$ICREA/IFAE, Barcelona, Spain,
$^{e}$Centro de Investigacion en Computacion - IPN, Mexico City, Mexico,
$^{f}$ECFM, Universidad Autonoma de Sinaloa, Culiac\'an, Mexico,
and 
$^{g}$Universit{\"a}t Bern, Bern, Switzerland.%
}} \noaffiliation
\vskip 0.25cm

%% file: acknowledgement.tex
%
We thank the staffs at Fermilab and collaborating institutions,
and acknowledge support from the
DOE and NSF (USA);
CEA and CNRS/IN2P3 (France);
FASI, Rosatom and RFBR (Russia);
CNPq, FAPERJ, FAPESP and FUNDUNESP (Brazil);
DAE and DST (India);
Colciencias (Colombia);
CONACyT (Mexico);
KRF and KOSEF (Korea);
CONICET and UBACyT (Argentina);
FOM (The Netherlands);
STFC and the Royal Society (United Kingdom);
MSMT and GACR (Czech Republic);
CRC Program and NSERC (Canada);
BMBF and DFG (Germany);
SFI (Ireland);
The Swedish Research Council (Sweden);
and
CAS and CNSF (China).

%% file: wz_2010_PLB.bbl
\begin{thebibliography}{99}

\bibitem{ssm} J.C.~Pati, A.~Salam, Phys. Rev. D {\bf 10}, 275 (1974) [Erratum ibid. D {\bf 11}, 703 (1975)];
R.N.~Mohapatra, J.C.~Pati, Phys. Rev. D {\bf 11}, 566 (1975); G.~Senjanovic, 
R.N.~Mohapatra, Phys. Rev. D {\bf 12}, 1502 (1975);
G.~Altarelli, B.~Mele, M.~Ruiz-Altaba, Z. Phys. C {\bf 45} (1989) 109 [Erratum ibid. C {\bf 47}, 676 (1990)].

\bibitem{extradim} H.~He $et~al.$, Phys. Rev. D {\bf 78}, 031701 (2008); A.~Belyaev, arXiv:0711.1919 [hep-ph] (2007); K.Agashe {\it et al.}, Phys. Rev. D {\bf 80}, 075007 (2009). 

\bibitem{litHig} M.~Perelstein, Prog. Part. Nucl. Phys. {\bf 58}, 247 (2007).

\bibitem{LSTC} E.~Eichten~and~K. Lane, Phys. Lett. B {\bf{669}}, 235 (2008); K.~Lane, Phys. Rev. D {\bf{60}}, 075007 (1999).

\bibitem{Th-Hagiw1} K.~Hagiwara, R. D.~Peccei, and D.~Zeppenfeld, Phys. Rev. B {\bf 282}, 253 (1987).

\bibitem{Th-Hagiw2} K.~Hagiwara, J.~Woodside, and D.~Zeppenfeld, Phys. Rev. D {\bf 41}, 2113(1990).

\bibitem{Coup-HISZ} K.~Hagiwara, S.~Ishihara, R.~Szalapski, and D.~Zeppenfeld,
Phys. Rev. D {\bf 48}, 2182 (1993); Phys. Lett. B {\bf 283}, 353 (1992).

\bibitem{CDF-CrossSec} A.~Abulencia {\it et al.}, CDF Collaboration, Phys. Rev. Lett. {\bf 98}, 161801 (2007).

\bibitem{D0RunIIaWZ}  V.~Abazov {\it et al.}, D0 Collaboration, Phys. Rev. D {\bf 76}, 111104 (2007).

\bibitem{LEP} The LEP Collaborations ALEPH, DELPHI, L3, OPAL, http://lepewwg.web.cern.ch/LEPEWWG/lepww/tgc/\\summer03/gc\_main2003.ps .

\bibitem{CDF-coup} T. Aaltonen {\it et al.}, CDF Collaboration, Phys. Rev. D {\bf 76}, 111103 (2007).

\bibitem{delta_r} The D0 detector uses a right-handed coordinate system with the $z$ axis 
pointing in the direction of the proton beam and the $y$ axis pointing upwards. The azimuthal angle  
$\phi$ is defined in the $xy$ plane measured from the $x$ axis. The pseudorapidity is defined as 
$\eta = -\rm{ln}[\rm{tan}(\theta/2)]$, where $\theta = {\rm{\arctan}}(\sqrt{x^2+y^2}/z)$. The transverse variable
is defined as projection onto the $x-y$ plane. The missing transverse energy is the 
imbalance of the momentum estimated from the calorimeter cells and reconstructed muons in the
$x-y$ plane.

\bibitem{run2det} V.M. Abazov {\it et al.}, D0 Collaboration, Nucl. Instrum. Methods Phys. Res. A {\bf 565}, 463 (2006).

\bibitem{pythia} T.~Sj\"{o}strand, S. Mrenna, and P. Skands, J.~High~Energy~Phys. {\bf 05}, 026 (2006); we used V6.419.

\bibitem{alpgen} M.~L.~Mangano $et~al.$, J. High Energy Phys. {\bf 07}, 1 (2003).

\bibitem{geant} {\sc GEANT} Detector Description and Simulation Tool, CERN Program Library Long Writeup W5013.

\bibitem{Baur} U.~Baur and E.~Berger, Phys. Rev. D {\bf 47}, 4889 (1993).

\bibitem{ttbar-cross-sec} P.~M.~Nadolsky $et~al.$, Phys. Rev. D {\bf 78}, 013004 (2008). 

\bibitem{zz-cross-sec} J.~M.~Campbell and R.~K.~Ellis, Phys. Rev. D {\bf 60}, 113006 (1999).

\bibitem{lhood} G.J.~Feldman and R.D.~Cousins, Phys. Rev. D {\bf 57}, 3873 (1998).

\bibitem{lumi} T.~Andeen $et~al.$, FERMILAB-TM-2365 (2007).

\end{thebibliography}
